\documentclass[conference]{IEEEtran}
\usepackage{amsmath,amsfonts,amssymb}
\usepackage{graphicx}
\usepackage{caption}
\usepackage{subcaption}
\usepackage{multirow}
\usepackage{algorithmic}
\usepackage[ruled]{algorithm}

\date{}

\include{Qcircuit}

\title{Software Pauli Tracking for Quantum Computation}
\author{\IEEEauthorblockN{Alexandru Paler$^1$ \qquad Simon Devitt$^2$
    \qquad Kae Nemoto$^2$ \qquad Ilia Polian$^1$}\\[-0.4cm]
  \IEEEauthorblockA{
  \begin{tabular}{c@{\hspace{2cm}}c}
    $^1$Faculty of Informatics and Mathematics & $^2$National Institute of Informatics \\
    University of Passau & 2-1-2 Hitotsubashi, Chiyoda-ku \\
    Innstr.~43, D-94032 Passau, Germany & Tokyo, Japan \\
    \{alexandru.paler$|$ilia.polian\}@uni-passau.de & \{devitt$|$nemoto\}@nii.ac.jp
  \end{tabular}
}}

\begin{document}
\maketitle

\begin{abstract}
The realisation of large-scale quantum computing is no longer simply a hardware question.  
The rapid development of quantum technology has resulted in dozens of control and programming problems that should be directed towards the classical computer science 
and engineering community.  One such problem is known as Pauli tracking.  Methods for implementing 
quantum algorithms that are compatible with crucial error correction technology utilise 
extensive quantum teleportation protocols.  These protocols are intrinsically probabilistic 
and result in correction operators that occur as byproducts of teleportation.  These 
byproduct operators do not need to be corrected in the quantum hardware itself.  Instead, 
byproduct operators are tracked through the circuit and output results \emph{reinterpreted}. 
This tracking is routinely ignored in quantum information as it is assumed that tracking algorithms will eventually be developed.  In this work we help fill this gap and present an 
algorithm for tracking byproduct operators through a quantum computation.  We formulate 
this work based on quantum gate sets that are compatible with all major forms of 
quantum error correction and demonstrate the completeness of the algorithm.
\end{abstract}

\section{Introduction}

Quantum computing promises exponential speed-up for a
number of relevant computational problems. Building a
scalable and reliable quantum computer is one of the
grand challenges of modern science. While
small-scale quantum computers are routinely being
fabricated and operated in the laboratory 
\cite{HA08,PLZY08,PMO09,BW08,P12,L12}, they can only serve as
feasibility studies, and fundamental breakthroughs will
be required before a truly practical quantum computer
can be built. As the size of computers 
within the reach of state-of-the-art technologies increases,
the focus of interest shifts from their basic physical
principles to structured design methodologies that will
allow to realise large-scale systems
\cite{CMS07,P12+,SM13,SWD11}.

A given technology is suited for construction of general-purpose
quantum computers if it supports a direct realisation of a
\emph{universal quantum gate set} which can implement
or approximate arbitrary functions \cite{NC00}. Moreover, today's quantum
systems exhibit high error rates and require effective
\emph{quantum error-correcting codes} (QECC) \cite{DMN13}.
Consequently, building a practical quantum computer
requires an universal gate set which can be implemented
in an error-corrected manner.

In this paper, we consider a class of quantum circuits
based on an universal gate set that consists of just two
types of operations: injection of specific quantum states
into the circuit and the controlled-not (CNOT) operation.
Using the technique of \emph{quantum teleportation},
state injections are mapped to \emph{rotational gates}
that together with the CNOT operation provide universality \cite{BK05+}.
The advantage of this gate set is that it can be seamlessly
integrated into very advanced QECC schemes, allowing 
for scalable, large-scale information processing \cite{DMN13,FM12}. However, 
as quantum teleportation is inherently
probabilistic the direction of qubit rotations is random.  This 
randomness can be corrected via a technique known as 
\emph{Pauli tracking}.  Pauli tracking operates by constructing a 
classical record of each teleportation result and  reinterprets 
later results during the computation.  This tracking means that 
we do not need to perform active quantum corrections because of the 
probabilistic nature of teleportation operations.  This technique is 
well known in the quantum information community and routinely 
ignored (referred to as working in the \emph{Pauli frame}).  However, 
to our knowledge, no details on the algorithm necessary to perform this 
tracking have been presented.


The contributions of this paper are as follows. First, we
present teleportation-based quantum computing in a
generic and algorithmic way accessible to the design
community.  There are many types of corrections that are combined 
to define the Pauli frame of a quantum computation.  The two most important 
arise from teleportation operations when performing arbitrary rotations and the 
second arises from QEC decoding operations.  Without loss of generality we will 
focus on the first.  By doing this, we completely
detach our description from a particular type of QECC; in fact, an
arbitrary QECC can be applied on top of the basic scheme with 
a minor adjustment to the algorithm.

Second, we introduce a new algorithm for \emph{Pauli
tracking}. This algorithm allows us to postpone corrections
until the end of computation where we adjust the output of the 
computation based on the current state of the Pauli frame.
The algorithm is completely classical and can be implemented
in software and run on the control computer rather than on
the quantum hardware. We formalise the algorithm and
prove its correctness. Experimental results show that Pauli
tracking is efficient.

The paper is organised as follows: In section \ref{sec:Quant} we introduce the basics of quantum 
computation.  Section \ref{sec:ft} details the compatible gate sets for fault-tolerant, 
error-corrected quantum computation and section \ref{sec:teleport} introduces teleportation-based quantum 
gates.  Finally, section \ref{sec:corr} illustrates the Pauli tracking algorithm and section \ref{sec:sims} 
presents several simulation results.



\section{Quantum Computing}
\label{sec:Quant}

Quantum circuits represent and manipulate information in
\emph{qubits} (quantum bits). While classical bits assume
either logic value 0 or 1, qubits may be in \emph{superposition}
of these two values. A single qubit has a \emph{quantum state}
$\ket{\psi}= (\alpha_0, \alpha_1)^T = \alpha_0\ket{0} +
\alpha_1\ket{1}$. Here, $\ket0 = (1, 0)^T$ and $\ket1 = (0, 1)^T$
are quantum analogons of classical logic values 0
and 1, respectively. $\alpha_0$ and $\alpha_1$ are complex
number called \emph{amplitudes} with $|\alpha_0|^2 +
|\alpha_1|^2 = 1$. $\ket0$ and $\ket1$ are orthonormal vectors
and form a basis of $\mathbb{C}^2$.

A state $(\alpha_0, \alpha_1)^T$ may
be modified by applying single-qubit \emph{quantum gates}.
Each quantum gate corresponds to a complex unitary matrix,
and gate function is given by multiplying that matrix with the
quantum state. Two single-qubit gates that are highly relevant
in the context of this paper are the $X$ and the $Z$ gate with
the following matrices:
\begin{align*}
X = \begin{pmatrix}
0 & 1 \\
1 & 0
\end{pmatrix} 
&&
Z = \begin{pmatrix}
1 & 0 \\
0 & -1
\end{pmatrix} 
\end{align*}
The application of $X$ to a state results in a \emph{bit flip}:
$X(\alpha_0, \alpha_1)^T = (\alpha_1, \alpha_0)^T$; $\ket0$ is
mapped to $\ket1$, and vice versa. The application of the $Z$
gate results in a \emph{phase flip}: $Z(\alpha_0, \alpha_1)^T
= (\alpha_0, -\alpha_1)^T$. Bit and phase flips are used for
modelling the effects of errors on the quantum state as qubit errors can 
be decomposed into combination of bit and/or phase flips. Further important single-qubit
quantum gates, in the context of a fully error-corrected system, are
\begin{align*}
H = \frac1{\sqrt2}\begin{pmatrix}
1 & 1 \\
1 & -1
\end{pmatrix} 
&&
P = \begin{pmatrix}
1 & 0 \\
0 & i
\end{pmatrix} 
&&
T = \begin{pmatrix}
1 & 0 \\
0 & e^{i\frac{\pi}{4}}
\end{pmatrix}, 
\end{align*}
where $T^2 = P$ and $P^2 = Z$.  

It is not possible to
directly read out the amplitudes of a qubits state. A
\emph{measurement} has to be performed instead. Quantum
measurement is defined with respect to a basis and yields one
of the basis vectors with a probability related to the amplitudes of the quantum 
state. Of importance in this work
are $Z$- and $X$-measurements. $Z$-measurement is defined
with respect to basis $(\ket0, \ket1)$. Applying a $Z$-measurement
to a qubit in state $\ket{\psi}= \alpha_0\ket{0} + \alpha_1\ket{1}$
yields $\ket0$ with probability $|\alpha_0|^2$ and $\ket1$ with
probability $|\alpha_1|^2$. Moreover, the state $\ket\psi$
\emph{collapses} into the measured state, that is, becomes either
$\ket0$ or $\ket1$. $X$-measurement is defined with respect to
the basis $(\ket+, \ket-)$, where $\ket{+} = \frac{1}{\sqrt2}(\ket{0} +
\ket{1})$ and $\ket{-}=\frac{1}{\sqrt2}(\ket{0}-\ket{1})$.

A multi-qubit circuit processes states represented by an exponential
number of amplitudes. The state of a circuit with $n$ qubits has $2^n$
amplitudes $\alpha_y$ with $y \in \{0, 1\}^n$ and $\sum_y|\alpha_y|^2
= 1$. For example, the state of a 2-qubit circuit is $\ket{\psi}= (\alpha_{00},
\alpha_{01}, \alpha_{10}, \alpha_{11})^T = \alpha_{00}\ket{00} +
\alpha_{01}\ket{01} + \alpha_{10}\ket{10} + \alpha_{11}\ket{11}$.
Here, $\ket{00} = (1, 0, 0, 0)^T$, $\ket{01} = (0, 1, 0, 0)^T$, $\ket{10} = (0, 0, 1, 0)^T$
and $\ket{11} = (0, 0, 0, 1)^T$ form a basis for $\mathbb{C}^4$. Measuring
multiple qubits of a circuit again results in one basis vector with the probability
given by the corresponding amplitude, $|\alpha_y|^2$.

Quantum gates may act on several qubits simultaneously. A
gate that acts on $n$ qubits is represented by a $2^n \times
2^n$ complex unitary matrix. One important two-qubit gate
is the \emph{controlled-not} $CNOT(c,t)$ gate, where the $c$ qubit 
conditionally flips the state of the $t$ qubits when set to $\ket{1}$. Moreover, it is
possible to represent a single-qubit gate (or, more generally,
a gate acting on less qubits than $n$) by using tensor product.
For example, the $2^2 \times 2^2$ matrix $H \otimes I$
(where $I$ is the identity matrix) applies the Hadamard gate
$H$ to the first qubit of a two-qubit circuit while leaving the
second qubit unchanged. The first qubit of the CNOT gate is called \emph{control qubit}
$c$ and the second is the \emph{target qubit} $t$. Below are the matrices of the
$CNOT(c=1,t=2)$ and the $H \otimes I$ operations.
\begin{align*}
CNOT \!=\! \begin{pmatrix}
1 & 0 & 0 & 0 \\
0 & 1 & 0 & 0 \\
0 & 0 & 0 & 1 \\
0 & 0 & 1 & 0
\end{pmatrix} 
&&
H \!\otimes\! I \!=\! \frac1{\sqrt2}\!\begin{pmatrix}
1 & 0 & 1 & 0 \\
0 & 1 & 0 & 1 \\
1 & 0 & -1 & 0 \\
0 & 1 & 0 & -1
\end{pmatrix} 
\end{align*}

A quantum circuit with $n$ qubits and $m$ gates $g_1, \ldots, g_m$
takes an \emph{input state} $\phi_0 \in \mathbb{C}^{2^n}$
and successively applies the transformations corresponding to each
gate: $\phi_1 = M_{g_1} \phi_0$; $\phi_2 = M_{g_2} \phi_1$;
and so forth, where $M_{g_i}$ is the $2^n \times 2^n$ matrix of
gate $g_i$. Selective qubit measurement may also be interspersed with these gates and at the 
end of computation the \emph{output state} $\phi_m$ of the circuit is 
completely measured. 

\section{Fault-Tolerant Gate Set}
\label{sec:ft}

It is important to distinguish between the generic mathematical model
of quantum computation and subsets of it that are suited for an
actual physical realisation. A number of technologies have been suggested
for implementing quantum computation \cite{DFSG08,YJG10,MLFY10,JMFMKLY10}. 
While every unitary
complex matrix qualifies as a quantum gate in the mathematical formalism,
most implementation technologies only allow a direct physical realisation of
relatively few gates. Therefore, \emph{universal gate sets}, that is, collections
of quantum gates that can represent or approximate an arbitrary quantum
circuit, are of interest This concept is similar to universal gate libraries in
digital circuit design, where each Boolean function can be mapped to a
circuit composed of, for instance, AND2 gates and inverters. One instance
of a universal quantum gate set is $\{CNOT, H, P, T\}$. A technology is
suitable for realisation of arbitrary quantum algorithms if it has a direct implementation
for at least one universal gate set.

A further key requirement for successful realisation of quantum
circuits is the ability to perform \emph{error correction} during
computation. States of actual quantum systems are inherently fragile
and are affected even by the slightest interaction with their environment.
Therefore, \emph{quantum error-correcting codes} (QECC) introduce substantial
redundancy to compensate for impact on the quantum state. For instance,
Shor's code \cite{S95} uses nine qubits to represent one \emph{encoded} qubit: the
qubit is first triplicated in order to detect and correct phase flips, and the
resulting qubit triplet is again triplicated to protect them against bit flips.  
It can be shown that this construction is sufficient to correct all errors
affecting any one of the nine physical qubits. We call the nine qubits used
for encoding \emph{physical qubits} and one error-corrected qubit
\emph{logical qubit}. The strength of a QECC can be quantified by how many 
physical qubit errors that have to occur before the logical qubit is corrupted. 
For Shor's code, a single error on any of the nine physical qubit is
tolerated; the probability of failure for the Shor's code is bounded by
the probability of two or more errors occurring on separate qubits. More advanced
codes can tolerate multiple errors and have lower probabilities
of failure at the expense of more physical qubits to encode a single logical qubit.

A \emph{fault-tolerant gate} for a given QECC acts directly on encoded
qubits and produces legal encodings with respect to that QECC at its
outputs. For example, a fault-tolerant implementation of a CNOT gate
for the Shor's code would take 18 physical qubits as inputs, interpret
nine of the qubits as the logical control qubit and the other nine qubits
as the logical target qubit, and produce 18 physical qubits that encode
the two logical qubits as its outputs. In self-checking design of classical
circuits, error-correcting codes with this property are called \emph{closed}
with respect to the gate's operation; for example, bi-residue codes are
closed under both addition and multiplication \cite{KK07}. The
closure property is advantageous because decoding and re-encoding
of codewords before and after operation are avoided. This advantage is
even more pronounced for quantum circuits with their extremely high
expected error rates. As a consequence, a practical universal gate set
should consist of fault-tolerant gates that allow circuit operation with
errors continuously taking place.

\section{Teleporation-based Quantum Computing}
\label{sec:teleport}

In this paper, we focus on a set of operations that can be
implemented in a fault-tolerant manner with respect to several
state-of-the-art QECC \cite{BK05+}.
This set consists of the CNOT gate and two \emph{state injection}
operations. State injection refers to initializing a qubit in one of the two following states 
$$\ket{A}=\frac{1}{\sqrt{2}}(\ket{0}+e^{i\pi/4}\ket{1})
\qquad
\ket{Y}=\frac{1}{\sqrt{2}}(\ket{0}+i\ket{1}).$$
Using these states and a technique called
\emph{quantum teleportation}, it is possible to obtain the
following three quantum gates:
\begin{align*}
R_x\left(\frac\pi4\right) = \frac1{\sqrt2}\begin{pmatrix}
1 & -i \\
-i & 1
\end{pmatrix} 
\end{align*}
\begin{align*}
R_z\left(\frac{\pi}{4}\right) = P =
\begin{pmatrix}
1 & 0 \\
0 & i
\end{pmatrix} 
&&
R_z\left(\frac{\pi}{8}\right) = T =
\begin{pmatrix}
1 & 0 \\
0 & e^{i\frac{\pi}{8}}
\end{pmatrix} 
\end{align*}
In the \emph{Bloch sphere representation} of a quantum state,
$R_x(\theta)$ and $R_z(\theta)$ stand for a rotation around
the $X$- and the $Z$-axis by angle $\theta$; see \cite{NC00} for
details. We use the following abbreviations for brevity: $R^4_x
:= R_x(\pi / 4); R^4_z := R_z(\pi / 4) \equiv P; R^8_z := R_z(\pi / 8)\equiv T$.
Using the relationship $H = R_z^4R_x^4R_z^4$, the complete
universal gate set $\{CNOT, H, P, T\}$ can be obtained based
on CNOT, state injection and quantum teleportation.  All these
gates are compatible with error-corrected, fault-tolerant computation \cite{BK05+,DSMN13}. 
However, quantum teleportation is
probabilistic itself and may require (classical) correction that will be tracked. This is
described in detail below.

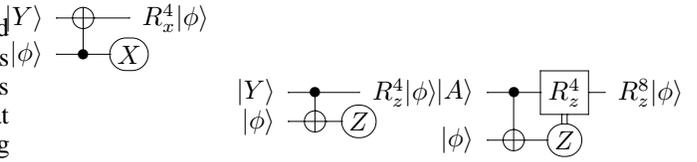
\begin{figure}
\centering
\Qcircuit @C=.6em @R=.35em {
\lstick{\ket{Y}} & \targ & \rstick{R_x^4\ket{\phi}} \qw \\
\lstick{\ket{\phi}} & \ctrl{-1} & \measure{X}
}\hspace{0.08\textwidth}
\label{fig:teleb}%
\Qcircuit @C=.6em @R=.35em {
\lstick{\ket{Y}} & \ctrl{1} & \rstick{R_z^4\ket{\phi}} \qw\\
\lstick{\ket{\phi}} & \targ & \measure{Z}
}\hspace{0.08\textwidth}
\Qcircuit @C=.6em @R=.35em {
\lstick{\ket{A}} & \ctrl{1} & \gate{R_z^4} & \rstick{R_z^8\ket{\phi}} \qw\\
\lstick{\ket{\phi}} & \targ & \measure{Z} \cwx
}
\caption{Teleportation circuits used for (a) $R_x^4$, (b) $R_z^4$, (c) $R_z^8$}
\label{fig:tele}
\end{figure}

The rotational gates $R^4_x$, $R^4_z$ and $R^8_z$ are
constructed by combining state injection with quantum teleportation.
Applying the three employed rotational gates to an arbitrary
state $\ket{\phi}$ by quantum teleportation is shown in
Fig.~\ref{fig:tele}. An auxilliary qubit is initialised in state $\ket{Y}$
or $\ket{A}$ (depending on the desired rotation), and a CNOT
gate is applied at the qubit that holds $\ket{\phi}$ and the
auxilliary qubit (the control and target qubits are denoted by
$\bullet$ and $\oplus$, respectively).
Finally, a measurement (either $X$ or $Z$)
is performed at the control output of the CNOT gate, indicated
in Fig.~\ref{fig:tele} by an encircled $X$ or $Z$. The effect at
the target output is shown in Fig. \ref{fig:tele}

\noindent{\bf Gate $R^4_x:$} \;
The $X$-measurement in circuit of Fig.~\ref{fig:tele}a yields
either $\ket{+}$ or $\ket{-}$. If the measurement result is
$\ket{+}$, then the desired rotation $R_x(\pi / 4)$ was executed
and the new state at the output of the circuit is correct. If the
measurement result is $\ket{-}$, the applied rotation was
$R_x(-\pi / 4)$, i.e., the direction of the rotation was wrong.
This is easily compensated by performing another rotation by
angle $\pi/2$, namely applying the gate $R_x(\pi/2) = X$. It is
easily checked that $XR_x(-\pi / 4) = R_x(\pi / 4)$. Consequently,
quantum teleportation must be followed by executing the $X$
gate at the obtained state if the measurement result is $\ket{-}$.
We call this \emph{$X$-correction}. Note that the decision whether
$X$-correction is required is based on classical information (a
measurement result) and can be taken by a classical computer.

\noindent{\bf Gate $R^4_z:$} \;
The two possible $Z$-measurement results from the circuit in
circuit of Fig.~\ref{fig:tele}b implementing $R_z^4$ are
$\ket{0}$ and $\ket{1}$. The state $\ket{0}$ indicates a correct
teleportation where the resulting state is
$\ket{\psi}=R_z(\pi/4)\ket{\phi}$. 
A measured state $\ket{1}$ is an indicator for the state
$\ket{\psi_f}=\alpha_1\ket{0}-i\alpha_0\ket{1}$ where the
input state was $\ket{\phi}=\alpha_0\ket{0} + \alpha_1\ket{1}$.
In order to obtain the correct state, a $Z$ operation followed
by the $X$ operation is applied, as it is easily verified that
$\ket{\psi}=XZ\ket{\psi_f}=\alpha_0\ket{0}+i\alpha_1\ket{1} = R^4_z\ket{\phi}$.
This operation is called $XZ$ correction.

\noindent{\bf Gate $R^8_z:$} \;
This gate is implemented in two stages (see Fig.~\ref{fig:tele}c).
The first teleporation maps state $\ket A$ to an intermediate state,
which is then given to the $R_z^4$ gate from Fig.~\ref{fig:tele}b
that also incorporates a teleportation. The following three measurement
outcomes have to be distinguished:
\begin{enumerate}
\item If the first measurement results in $\ket{0}$, the intermediate
state is the correct result already. No correction is required, and
the second rotation (including the corresponding measurement) does
not have to be applied.
\item If the first measurement results in $\ket{1}$ and the input
state was $\ket{\phi}=\alpha_0\ket{0} + \alpha_1\ket{1}$, the
calculated state is $\ket{\psi_{f1}}=XR_z^{4\dagger}\ket{\phi}
=\alpha_0\ket{1}+e^{-i\pi/4}\alpha_1\ket{0}$. This state will be
used as an input for the $\pi/2$ correctional rotation ($R_z^4$
from Fig.~\ref{fig:tele}b). If the second measurement returns
$\ket{1}$, then the correction succeeded, and no further
corrections are necessary: $\ket{\psi}=\alpha_0\ket{0} +
e^{i\pi/4}\alpha_1\ket{1}=R^8_z\ket{\phi}$.
\item If the first measurement returns $\ket{1}$, and the
second measurement yields $\ket{0}$,  then the $R_z^4$
correction will produce state $\ket{\psi_{f2}} = i\alpha_0\ket{0}
+ e^{-i\pi/4}\alpha_1\ket{0}$. Then, as seen above, the
$XZ$ correction leads to $\ket{\psi}=XZ\ket{\psi_{f2}} =
\alpha_0\ket{0}+e^{i\pi/4}\alpha_1\ket{1}=R_x^8\ket{\phi}.$
\end{enumerate}

In summary, teleporations are probabilistic and either $X$
or $XZ$ corrections may be required depending on the outcomes
of the measurement. It is important to understand that this
non-determinism is \emph{not} due to errors but is inherent
to teleportation-based quantum computing. Algorithm
\ref{alg:tele} summarises the complete computation procedure
incorporating all the required corrections in detail. The algorithm
assumes a circuit that has already been mapped to the universal
gates set consisting of the CNOT gate and the three considered
rotational gates. Note that the only operations applied are
state injections and CNOT gates, and that these operations are
compatible with standard fault-tolerant error correction. 

The computation contiunuously requests new qubits and
abandons the old ones, such that the total number of used logical
qubits is $n$ or $n + 1$ at any given time. In many relevant
implementation technologies, hardware for each
abandoned qubit can be reused for the newly requested ones.
For example, if a new qubit is introduced for injecting the $\ket Y$
state in order to implement the $R^4_x$ gate, the old qubit is
no longer required after the $X$-measurement, and it can be
used for implementing further rotational gates.

\renewcommand{\algorithmicrequire}{\textbf{Input:}}
\renewcommand{\algorithmicensure}{\textbf{Output:}}
\begin{algorithm}[t]
  \caption{Teleportation-based quantum computation}
  \label{alg:tele}
  \begin{algorithmic}[1]
    \REQUIRE{$n$-qubit quantum circuit with $m$ gates
      $g_1, \ldots, g_m \in \{CNOT, R^4_x, R^4_z, R^8_x\}$, input state $\phi_0$}
    \ENSURE{Output state $\phi_m$}
    \FOR{$i := 1$ \TO $m$}
      \IF{$g_i$ is a CNOT gate}
        \STATE // Apply CNOT to current state
        \STATE $\phi_i := M_g \phi_{i + 1};$
      \ELSIF{$g_i$ is a $R^4_x$ gate on qubit $k$}
        \STATE Introduce new qubit $l$; inject state $\ket Y$ on $l$;
        \STATE Perform CNOT($k$, $l$); $X$-measurement on qubit $k$;
        \IF{measurement result is $\ket{+}$}
          \STATE apply $X$-correction on qubit $l$;
        \ENDIF
        \STATE Replace qubit $k$ in $\phi_{i - 1}$ by qubit $l$ to obtain $\phi_i$;
      \ELSIF{$g_i$ is a $R^4_z$ gate}
        \STATE Introduce new qubit $l$; inject state $\ket Y$ on $l$;
        \STATE Perform CNOT ($l$, $k$); $Z$-measurement on $k$;
        \IF{measurement result is $\ket{0}$}
          \STATE Apply $XZ$-correction on qubit $l$;
        \ENDIF
        \STATE Replace qubit $k$ in $\phi_{i - 1}$ by qubit $l$ to obtain $\phi_i$;
      \ELSIF{$g_i$ is a $R^8_z$ gate}
        \STATE Introduce new qubit $l$; inject state $\ket A$ on $l$;
        \STATE Perform CNOT($l$, $k$); $Z$-measurement on $k$;
        \IF{measurement result is $\ket{0}$}
          \STATE Introduce new qubit $l'$; inject state $\ket Y$ on $l$;
          \STATE Perform CNOT($l'$,$l$); $Z$-measurement on $l$;
          \IF{measurement result is $\ket{0}$}
            \STATE Apply $XZ$-correction on qubit $l'$;
          \ENDIF
          \STATE Replace qubit $k$ in $\phi_{i - 1}$ by qubit $l'$ to obtain $\phi_i$;
        \ELSE  
          \STATE Replace qubit $k$ in $\phi_{i - 1}$ by qubit $l$ to obtain $\phi_i$;
        \ENDIF
      \ENDIF
    \ENDFOR
    \RETURN $\phi_m$;
  \end{algorithmic}
\end{algorithm}

\section{Pauli Tracking Algorithm}
\label{sec:corr}

Teleportation-based quantum computing suffers from the necessity
to conditionally perform corrections based on the measurement
results. Applying correction in real time, immediately after the
rotational operation, such as in Algorithm
\ref{alg:tele}, results in significant interaction between quantum
hardware and the classical control computer. This may have a
detrimental impact on the speed of computation and is ultimately unnecessary. In this
section, we demonstrate how applying corrections can be
postponed to the end of calculation without losing accuracy. For
this purpose, teleportation-based quantum computation method
is modified as follows.

The considered circuits still consist of CNOT gates and the three
types of rotational gates implemented by state injection and
quantum teleportation. Measurements are still performed during
quantum teleportation, however their outcomes are stored in a
variable rather than used for immediate correction. For each
rotational gate $g_i$, variable $b_i$ holds the result of the
measurement. Note that $b_i \in \{\ket+, \ket-\}$ if $g_i$ is
a $R^4_x$ gate, $b_i \in \{\ket0, \ket1\}$ if $g_i$ is
a $R^4_z$ gate, $b_i \in \{\ket{00}, \ket{01}, \ket{10},
\ket{11}\}$ if $g_i$ is a $R^8_z$ gate, where pairs of values
refer to the outcomes of two consecutive measurements.

We derive an algorithm that calculates, for a given combination
of $b_i$ values, the vector of \emph{equivalent output correction
statuses} $S = (s_1, \ldots, s_n)$. For qubit $k$,
$s_k$ assumes one of four values that indicate the
required corrections: $I$ (no correction), $X$ ($X$-correction,
i.e., a bit flip), $Z$ ($Z$-correction, or a phase flip), and $XZ$
(both $X$- and $Z$-correction). The values in $S$ are
calculated such that running the teleportation-based quantum
computing (Algorithm \ref{alg:tele}) \emph{without applying
corrections in Lines 8--10, 15--17 and 22--31} and applying
the correction in $S$ to the obtained output state is
equivalent to teleportation-based quantum computing with
immediate correction.

$S$ is calculated by propagating (\emph{tracking}) the
correction status $(s_1, \ldots, s_n)$ through the circuit. The
calculation is applied \emph{after} the teleportation-based
quantum computation took place and the measurement
results $b_i$ associated with all rotational gates $g_i$ are
available. Each $s_k$ is initialised to $I$ (no correction).
Then, the gates are considered in their regular order $g_1,
\ldots, g_m$. If $g_i$ is a rotational gate on qubit $k$, its
$b_i$ is consulted to decide whether a correction is needed
and the $s_k$ is updated (the correction status is
propagated). The propagated correction status shows up
at the inputs of subsequent rotational and CNOT gates
and must be taken into account when calculating the
correction status at the output of that gates. We introduce
the \emph{correction status tracking function} $\tau$
that formalises the propagation.

There are two versions of $\tau$: one for CNOT gates
and for rotational gates. CNOT gates do not employ
teleportation and therefore require no corrections;
however, corrections that originated from rotational
gates may show up at the inputs of the CNOT gate
and have to be propagated to its outputs. Let $c$
and $t$ be the control and the target qubit of the
CNOT gate, and let $s_c^{\rm in}$ and $s_t^{\rm
in}$ be the correction statuses at the inputs of these
qubits, respectively. Then, $\tau(s_c^{\rm in},
s_t^{\rm in})$ produces a pair of correction statuses
$(s_c^{\rm out}, s_t^{\rm out})$ at the outputs
of the CNOT gates by the following calculation:
\begin{eqnarray}
\label{eq:tauc}
s_c^{\rm out} & = & \left\{\begin{array}{lcl}
s_c^{\rm in} & \mbox{ if } & s_t^{\rm in} \in \{I, X\}\\
s_c^{\rm in} \oplus Z & \mbox{ if } & s_t^{\rm in} \in \{Z, XZ\}
\end{array}\right.\\
\label{eq:taut}
s_t^{\rm out} & = & \left\{\begin{array}{lcl}
s_t^{\rm in} & \mbox{ if } & s_c^{\rm in} \in \{I, Z\}\\
s_t^{\rm in} \oplus X & \mbox{ if } & s_c^{\rm in} \in \{X, XZ\}
\end{array}\right.
\end{eqnarray}
Here, $s \oplus Z$ and $s \oplus X$ are flipping the status of the
respective correction in $s$:
$$\begin{array}{ccc|ccc}
s & s \oplus Z & s \oplus X & s & s \oplus Z & s \oplus X \\
\hline
I & Z & X & X & XZ & I \\
Z & I & XZ & XZ & X & Z
\end{array}$$

\begin{table}[t]
\centering
\caption{Correction status tracking $\tau$ for rotational gates}
\label{tab:tau}
\begin{tabular}{c@{\qquad}|@{\qquad}c}
\begin{tabular}{c|c|c|c}
$g_i$ & $s_k^{\rm in}$ & $b_i$ & $s_k^{\rm out}$ \\
\hline
\multirow{8}{*}{$R_x^4$}
& \multirow{2}{*}{$I$} & $\ket{+}$ & $I$\\
					        && $\ket{-}$ & $X$\\
\cline{2-4}
& \multirow{2}{*}{$Z$} & $\ket{+}$ & $X$\\
&& $\ket{-}$ & $I$\\
\cline{2-4}
 & \multirow{2}{*}{$X$} & $\ket{+}$ & $X$\\
      					     && $\ket{-}$ & $Z$\\
\cline{2-4}
 & \multirow{2}{*}{$XZ$} & $\ket{+}$ & $I$\\
      					     && $\ket{-}$ & $Z$\\
\hline
\multirow{8}{*}{$R_z^4$} & \multirow{2}{*}{$I$} & $\ket{0}$ & $I$\\
					        && $\ket{1}$ & $XZ$\\
\cline{2-4}
 & \multirow{2}{*}{$Z$} & $\ket{0}$ & $Z$\\
      					     && $\ket{1}$ & $X$\\
\cline{2-4}
	& \multirow{2}{*}{$X$} & $\ket{0}$ & $XZ$\\
      					     && $\ket{1}$ & $I$\\
\cline{2-4}
	& \multirow{2}{*}{$XZ$} & $\ket{0}$ & $X$\\
      					     && $\ket{1}$ & $X$\\
\end{tabular}
&
\begin{tabular}{c|c|c|c}
$g_i$ & $s_k^{\rm in}$ & $b_i$ & $s_k^{\rm out}$ \\
\hline
\multirow{13}{*}{$R_z^8$} & \multirow{3}{*}{$I$} & $\ket{0*}$ & $I$\\
					        && $\ket{10}$ & $XZ$\\
					        && $\ket{11}$ & $I$\\
\cline{2-4}
& \multirow{3}{*}{$Z$} & $\ket{0*}$ & $Z$\\
      					     && $\ket{10}$ & $X$\\
      					     && $\ket{11}$ & $Z$\\
\cline{2-4}
& \multirow{3}{*}{$X$} & $\ket{00}$ & $XZ$\\
      					     && $\ket{01}$ & $I$\\
						&& $\ket{1*}$ & $I$\\
\cline{2-4}
& \multirow{4}{*}{$XZ$} & $\ket{00}$ & $X$\\
      					     && $\ket{01}$ & $Z$\\
						&& $\ket{10}$ & $X$\\
						&& $\ket{11}$ & $Z$
\\\multicolumn{4}{c}{}
\\\multicolumn{4}{c}{}
\\\multicolumn{4}{c}{}
\end{tabular}
\end{tabular}
\end{table}

For a rotational gate $g_i$ on qubit $k$, the $\tau$ function takes the
pre-stored measurement result $b_i$ and the correction status $s_k^{\rm
in}$ at its input and calculates the correction status $s_k^{\rm out}$ at its
output. The values calculated by $\tau$ for the three types of rotational
gates considered are given in Table \ref{tab:tau}. 

Algorithm \ref{alg:track} summarises the Pauli tracking procedure.

\begin{algorithm}[t]
  \caption{Pauli tracking}
  \label{alg:track}
  \begin{algorithmic}[1]
    \REQUIRE{$n$-qubit quantum circuit with $m$ gates
      $g_1, \ldots, g_m \in \{CNOT, R^4_x, R^4_z, R^8_x\}$, measurement results
      $b_i$ for every rotational gate $g_i$}
    \ENSURE{Equivalent output correction status $S = (s_1, \ldots, s_n)$}
    \STATE $s_1 := s_2 := \cdots := s_n := I$;
    \FOR{$i := 1$ \TO $m$}
      \IF{$g_i$ is a CNOT gate with control/target qubits $c / t$}
        \STATE $(s_c, s_t) := \tau(s_c, s_t);$  \hspace{1cm} // Use Eqs.~\ref{eq:tauc}, \ref{eq:taut}
      \ELSIF{$g_i$ is a rotational gate on qubit $k$}
        \STATE $s_k := \tau(s_k, b_i)$; \hspace{1.64cm} // Use Table \ref{tab:tau}
      \ENDIF
    \ENDFOR
    \RETURN $S = (s_1, \ldots, s_n)$;
  \end{algorithmic}
\end{algorithm}

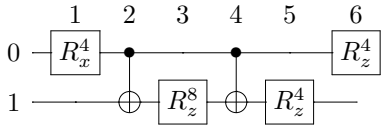
\begin{figure}[tb]
\centering
\hspace{0.1\textwidth}
\Qcircuit @C=.7em @R=.2em {
  & & 1 & 2 & 3 & 4 & 5 & 6\\
\\\\
0 && \gate{R_x^4} & \ctrl{1} & \qw & \ctrl{1} & \qw & \gate{R_z^4}\\
1 && \qw & \targ & \gate{R_z^8} & \targ & \gate{R_z^4} & \qw
}
\hspace{0.1\textwidth}
\caption{Quantum circuit implemented using fault-tolerant gates}
\label{fig:ex1}%
\end{figure}

\noindent{\bf Example:}\; Consider the two-qubit circuit in Fig.~\ref{fig:ex1}.
Assume that teleportation-based quantum computing has been done
without applying corrections. The recorded measurement results $b_i$
and the calculated correction statuses $S = (s_1, s_2)$ are shown in
the following table.

\begin{tabular}[t]{l|ccccccc}
$i$ &  & 1 & 2 & 3 & 4 & 5 & 6\\
\hline
$g_i$ && $R^4_x$ & CNOT & $R^8_z$ & CNOT & $R^4_z$ & $R^4_z$ \\
$b_i$ && $\ket+$ & n/a & $\ket{10}$ & n/a & $\ket0$ & $\ket1$ \\
\hline
$s_1$ & $I$ & $I$ & $I$ & $I$ & $Z$ & $Z$ & $X$\\
$s_2$ & $I$ & $I$ & $I$ & $XZ$ & $XZ$ & $X$ & $X$
\end{tabular}

Initially, $s_1$ and $s_2$ are set to $I$ (no correction required). Since
$b_1 = \ket+$ is measured for gate $g_1$, no correction is required and
$s_1$ remains $I$. The CNOT gate $g_2$ does not introduce new
corrections. The measurement outcome $b_3 = \ket{10}$ of gate $g_3$
necessitates the $XZ$-correction, as can be seen in Table \ref{tab:tau}.
Since the target input (qubit 2) of the CNOT gate $g_4$ includes $Z$,
the correction status at the control qubit is determined, using
Eq.~\ref{eq:taut}, as $s_1 = s_1 \oplus Z = I \oplus Z = Z$, while
$s_2$ remains unchanged. $b_5 = \ket0$ leads to $s_2 = X$ and
$b_6 = \ket1$ leads to $s_1 = X$. As a result, $X$-corrections must
be applied at both outputs of the circuit. \hfill$\Box$

The correctness of the tracking algorithm is now formally proven. The
following two lemmas formulate the validity of the $\tau$ function for
the individual gates. They can be verified for every combination of
inputs for $\tau$. Instead of providing the complete proof, we
quote the calculation for one specific input combination, whereas
the derivations for other combinations are similar.

\noindent\textbf{Lemma 1}\;
Let $g_i$ be a CNOT gate with correction status $s_c^{\rm in}$ at its
control and $s_t^{\rm in}$ at its target input and
$(s_c^{\rm out}, s_t^{\rm out}) = \tau(s_c^{\rm in}, s_t^{\rm in})$.
Then, performing the $s_c^{\rm in}$-correction at the control input
and the $s_t^{\rm in}$-correction at the target input followed by
application of CNOT is the same function as applying the CNOT gate
first and performing $s_c^{\rm out}$-correction at the control output
and the $s_t^{\rm out}$-correction at the target output.

\noindent\textbf{Proof} for $s_c^{\rm in} = X, s_t^{\rm in} = I$:\;
According to Eqs.~\ref{eq:tauc}, \ref{eq:taut}, $(s_c^{\rm out},
s_t^{\rm out}) = \tau(X, I) = (X, X)$. Without loss of generality,
assume that the control and target qubit of the CNOT gate are
qubit 1 and 2 respectively. Then, the $X$-correction at the control
qubit is described by matrix $X_1 = X \otimes I$ and the
$X$-correction at the target qubit is described by matrix $X_2 =
I \otimes X$. Performing the corrections first followed by the
CNOT operation corresponds to the matrix $CNOT \cdot X_1 \cdot I$
while the CNOT operation followed by the two corrections are
described by the matrix $X_1 \cdot X_2 \cdot CNOT$. 

The equivalence
of these matrices is shown below:
$$\begin{array}{l}
CNOT \cdot X_1 \\[0.1cm]
=\;\scriptstyle
\left(\begin{array}{cccc}
1&0&0&0\\
0&1&0&0\\
0&0&0&1\\
0&0&1&0
\end{array}\right)
\cdot\left(\begin{array}{cccc}
1&0&0&0\\
0&1&0&0\\
0&0&0&1\\
0&0&1&0
\end{array}\right)
= \left(\begin{array}{cccc}
0&0&1&0\\
0&0&0&1\\
0&1&0&0\\
1&0&0&0
\end{array}\right)\\[0.7cm]
=\;\scriptstyle
\left(\begin{array}{cccc}
0&0&1&0\\
0&0&0&1\\
1&0&0&0\\
0&1&0&0
\end{array}\right)
\cdot\left(\begin{array}{cccc}
0&1&0&0\\
1&0&0&0\\
0&0&0&1\\
0&0&1&0
\end{array}\right)
\cdot \left(\begin{array}{cccc}
1&0&0&0\\
0&1&0&0\\
0&0&0&1\\
0&0&1&0
\end{array}\right)\\[0.3cm]
=\;X_1 \cdot X_2 \cdot CNOT
\end{array}$$

All other combinations can be calculated similarly. \hfill$\Box$

\noindent\textbf{Lemma 2}\;
Let $g$ be a rotational gate, $s^{\rm in}$ the correction status
at its input, $b$ the outcome of the associated measurement and
$s^{out} = \tau(s^{\rm in}, b)$ the tracked correction status at
its output according to Table \ref{tab:tau}. Then, performing the
$s^{\rm in}$ correction, applying $g$ and perforimng, if needed,
correction according to $b$, yields the equivalent state as applying
$g$ first and performing $s^{\rm out}$-correction.

\noindent\textbf{Proof} for gate $R^4_z$, $b = \ket1$ und
$s^{\rm in} = Z$:\; Since $b = \ket1$, regular teleportation-based
computing requires an $XZ$-correction, such that the following four
operations are applied to the state: $Z$ for $s^{\rm in}$-correction;
$R^4_z$ for gate functionality, and $XZ$ for the correction of the
wrong rotation. From Table \ref{tab:tau}, $s^{\rm out} = \tau(Z,
\ket1) = X$ for the gate in question. The equivalence stated in
the lemma is verified by
$$\begin{array}{l}
XZR^4_zZ 
\scriptstyle
\left(\begin{array}{cc}
0&1\\
1&0
\end{array}\right)
\cdot
\left(\begin{array}{cc}
1&0\\
0&-1
\end{array}\right)
\cdot
\left(\begin{array}{cc}
1&0\\
0&i
\end{array}\right)
\cdot
\left(\begin{array}{cc}
1&0\\
0&-1
\end{array}\right)\\[0.7cm]
=\;\left(\begin{array}{cc}
0&i\\
1&0
\end{array}\right)
= \left(\begin{array}{cc}
0&1\\
1&0
\end{array}\right)
\cdot
\left(\begin{array}{cc}
1&0\\
0&i
\end{array}\right)
=XR^2_z
\end{array}$$

Other cases are checked similarly.\hfill$\Box$

Inductively applying the two lemmas to all gates in the circuit
leads to the validity of the following theorem.

\noindent\textbf{Theorem 1}\;
Applying corrections calculated by Algorithm \ref{alg:track} on the
state obtained by Algorithm \ref{alg:tele} without performing
immediate corrections results in the same state as the state
obtained by Algorithm \ref{alg:tele} when all corrections are
performed immediately.\hfill$\Box$



\begin{table}
\centering
\caption{Run-times $RT$ (in seconds)  of the Pauli tracking algorithm for circuits with $n$ qubits and $m$ quantum gates}
\label{tab:expres}
\begin{tabular}{rrr|rrr|rrr}
$n$ & $m$ & $RT$ [s] & $n$ & $m$ & $RT$ [s] & $n$ & $m$ & $RT$ [s] \\
\hline
100 & 1000 & 0 & 1100 & 1000 & 0.011 & 5100 & 1000 & 0.052 \\
100 & 5000 & 0.002 & 1100 & 5000 & 0.069 & 5100 & 5000 & 0.309 \\
100 & 10000 & 0.004 & 1100 & 10000 & 0.128 & 5100 & 10000 & 0.709 \\
100 & 20000 & 0.011 & 1100 & 20000 & 0.300 & 5100 & 20000 & 1.930 \\
100 & 50000 & 0.030 & 1100 & 50000 & 0.680 & 5100 & 50000 & 5.435\\
\end{tabular}
\end{table}

\section{Simulation Results}
\label{sec:sims}

We implemented the Pauli tracking algorithm and applied it
to a number of randomly generated quantum circuits with $n$
logical qubits and $m$ gates from the considered gate set.
The results for 100, 1100 and 5100 qubits are shown
in Table \ref{tab:expres}. $n = 100$ is indicative of largest
quantum circuits within reach of today's state-of-the-art
technology and $n = 1100$ and $n = 5100$ are the expected
sizes of quantum
computers within a decade. It can be seen that Pauli tracking
is fast and all calculations can be performed within a few
seconds for all cases.

The expected number of corrections
without Pauli tracking is $0.5\cdot m_4 + 0.75 m_8 \approx m$,
where $m_4$ is the number of gates $R^4_x$ and $R^4_z$
which require a correction with a probability 0.5 and $m_8$ is
the number of gates $R^8_z$ which may require one or two
corrections with the expected number of corrections equal to
0.75. The expected number of corrections with Pauli tracking is
bounded by $n$, because corrections have to be performed
on each output $k$ with $s_k \neq I$. As most relevant
circuits have far more gates than qubits, Pauli tracking
substantially reduces the overall effort for corrections.




\section{Conclusions}
\vspace*{-0.2cm}
We have presented an algorithm that can be used to perform Pauli tracking 
on quantum circuits compatible with all major classes of QECC.  This result helps fill 
an important gap in the classical control software needed for large-scale quantum 
computation.  Pauli tracking is instrumental for both error correction and for teleportation 
based protocols and this algorithm is easily adjustable to 
incorporate the required tracking for a specific implementation of quantum 
error correction.  Future work will be focused on adapting this algorithm to popular error 
correction techniques such as Topological codes \cite{FM12,RHG07} which requires 
even more intensive Pauli tracking due to the enormous number of teleportation 
gates necessary to perform a fully fault-tolerant, error corrected computation.


\vspace*{-0.2cm}

\begin{thebibliography}{1}
\expandafter\ifx\csname natexlab\endcsname\relax\def\natexlab#1{#1}\fi
\expandafter\ifx\csname url\endcsname\relax
  \def\url#1{\texttt{#1}}\fi
\expandafter\ifx\csname urlprefix\endcsname\relax\def\urlprefix{URL }\fi

\bibitem{HA08}
Hanson, R. \& Awschalom, D.
\newblock {Coherent manipulation of single spins in semiconductors}.
\newblock \emph{Nature (London)} \textbf{453}, 1043--1049 (2008).

\bibitem{PLZY08}
Press, D., Ladd, T.~D., Zhang, B. \& Yamamoto, Y.
\newblock {Complete quantum control of a single quantum dot spin using
  ultrafast optical pulses}.
\newblock \emph{Nature (London)} \textbf{456}, 218--221 (2008).

\bibitem{PMO09}
Politi, A., Matthews, J. \& O'Brien, J.
\newblock {Shor's quantum factoring algorithm on a photonic chip}.
\newblock \emph{Science} \textbf{325}, 1221 (2009).

\bibitem{BW08}
Blatt, R. \& Wineland, D.
\newblock {Entangled states of trapped atomic ions}.
\newblock \emph{Nature (London)} \textbf{453}, 1008--1015 (2008).

\bibitem{P12}
Pla, J. \emph{et~al.}
\newblock {A single-atom electron spin qubit in Silicon}.
\newblock \emph{Nature (London)} \textbf{489}, 541--545 (2012).

\bibitem{L12}
Lucero, E. \emph{et~al.}
\newblock {Computing prime factors with a Josephson phase qubit quantum
  processor}.
\newblock \emph{Nature Physics} \textbf{8}, 719--723 (2012).

\bibitem{CMS07}
Cheung, D., Maslov, D. \& Severini, S.
\newblock {Translationt echniques between quantum circuit architectures}.
\newblock \emph{Workshop on Quantum Information} (2013).

\bibitem{P12+}
Paler, A. \emph{et ~al.}
\newblock {Synthesis of topological quantum circuits}.
\newblock \emph{NANOARCH}, (2012).

\bibitem{SM13}
Saeedi, M. \& Markov, I.
\newblock {Synthesis and optimization of reversible circuits---a survey}.
\newblock \emph{ACM Computing Surveys} \textbf{45}, 21--34 (2013).

\bibitem{SWD11}
Saeedi, M., Wille, R. \& Drechsler, R.
\newblock {Synthesis of quantum circuits for linear nearest neighbor architectures}.
\newblock \emph{Quantum Information Processing} \textbf{10}, 355--377 (2011).

\bibitem{NC00}
M.A. Nielsen and I.L. Chuang.
\newblock {\em {Quantum Computation and Information}}.
\newblock Cambridge University Press, second edition, 2000.

\bibitem{DMN13}
Lucero, S.J., Munro, W.J. \& Nemoto, K.
\newblock {Quantum Error Correction for Beginners}.
\newblock \emph{Rep. Prog. Phys} \textbf{76}, 076001 (2013).

\bibitem{FM12}
Fowler, A., Mariantoni, M., Martinis, J. \& Cleland, A.
\newblock {Surface Codes, Towards practical large-scale quantum computation}.
\newblock \emph{Phys. Rev. A.} \textbf{86}, 032324 (2012).

\bibitem{DFSG08}
Devitt, S. \emph{et~al.}
\newblock {Architectural design for a topological cluster state quantum
  computer}.
\newblock \emph{New. J. Phys.} \textbf{11}, 083032 (2009).

\bibitem{YJG10}
Yao, N. \emph{et~al.}
\newblock {Scalable Architecture for a Room Temperature Solid-State Quantum
  Information Processor}.
\newblock \emph{Nature Communications} \textbf{3}, 800 (2012).

\bibitem{MLFY10}
Meter, R.~V., Ladd, T., Fowler, A. \& Yamamoto, Y.
\newblock {Distributed Quantum Computation Architecture Using Semiconductor
  Nonophotonics}.
\newblock \emph{Int. J. Quant. Inf.} \textbf{8}, 295--323 (2010).

\bibitem{JMFMKLY10}
Jones, N.~C. \emph{et~al.}
\newblock {A Layered Architecture for Quantum Computing Using Quantum Dots}.
\newblock \emph{Phys. Rev. X.} \textbf{2}, 031007 (2012{\natexlab{a}}).

\bibitem{S95}
Shor, P.W.
\newblock {A Scheme for reducing decoherence in quantum computer memory}.
\newblock \emph{Phys. Rev. A.} \textbf{52}, R2493 (1995).

\bibitem{KK07}
Koren, I. \& Krishna, C.~M.
\newblock {\em {Fault-Tolerant Systems}}.
\newblock Morgan-Kaufman Publishers, San Francisco, 2007.

\bibitem{BK05+}
Bravyi, S. \& Kitaev, A.
\newblock {Universal quantum computation with ideal Clifford gates and noisy
  ancillas}.
\newblock \emph{Phys. Rev. A.} \textbf{71}, 022316 (2005).

\bibitem{DSMN13}
Devitt, S.J., Stephens, A.M. Munro, W.J., \& Nemoto, K.
\newblock {Requirements for fault-tolerant factoring on an atom-optics quantum computer}.
\newblock \emph{Nature (communications)} In Press, (2013).

\bibitem{RHG07}
Raussendorf, R., Harrington, J. \& Goyal, K.
\newblock {Topological fault-tolerance in cluster state quantum computation}.
\newblock \emph{New J. Phys.} \textbf{9}, 199 (2007).

\end{thebibliography}
\end{document}